\documentstyle[12pt,lathuile]{article}
\def\vev#1{\left\langle #1\right\rangle}
\def\Im{\mathop{\mbox{Im}}}
\def\Re{\mathop{\mbox{Re}}}

\def\eoe{$\varepsilon'/\varepsilon$~}
\def\ratio{\eoe}
\def\eps{$\varepsilon$~}
\def\be{\begin{equation}}
\def\ee{\end{equation}}
\def\bea{\begin{eqnarray}}
\def\eea{\end{eqnarray}}
\def\Journal#1#2#3#4{{#1} {\bf #2}, #3 (#4)}

\def\NPB{{\em Nucl. Phys.} B}
\def\PLB{{\em Phys. Lett.}  B}
\def\PRL{\em Phys. Rev. Lett.}
\def\PRD{{\em Phys. Rev.} D}

\begin{document}
\title{ 
FINAL STATE INTERACTIONS IN \eoe
}
\author{
Stefano Bertolini       \\
{\em INFN and SISSA, 
Via Beirut 4, I-34013 Trieste, Italy} \\
{\em E-mail: bertolin@he.sissa.it}
}
\maketitle
\baselineskip=14.5pt
\begin{abstract}
I shortly review the present status of the
theoretical estimates of \eoe.
I consider a few aspects of the theoretical calculations
which may be relevant in understanding the present experimental
results. In particular, I discuss the role of final state interactions 
and in general of non-factorizable contributions for the
explanation of the $\Delta I = 1/2$ selection rule in kaon decays
and \eoe. Lacking reliable lattice calculations, the $1/N$ expansion
and phenomenological approaches may help in understanding 
correlations among theoretical effects and the experimental data.
The same dynamics which underlies the CP conserving selection rule
may drive \eoe in the range of the recent experimental measurements.
\end{abstract}
\baselineskip=17pt
\newpage

The results announced during the last year by the KTeV\cite{KTeV} 
and NA48\cite{NA48} collaborations
have marked a great experimental achievement,
establishing 35 years after the discovery of CP violation
in the neutral kaon system\cite{Christenson}
the existence of a much smaller violation acting directly in the
decays. 

While the Standard Model (SM) of strong and electroweak interactions
provides an economical and elegant understanding
of indirect~($\varepsilon$) and direct~($\varepsilon'$) 
CP violation in term of a single phase,
the detailed calculation of the size of these effects 
implies mastering strong interactions at a scale 
where perturbative methods break down. In addition, 
CP violation in $K\to\pi\pi$ decays
is the result of a destructive interference between
two sets of contributions,
which may inflate up to an order of magnitude the uncertainties
on the hadronic matrix elements of the effective 
four-quark operators.
All that makes predicting \eoe a complex and subtle theoretical
challenge\cite{review}.

\begin{figure}[t]
\vspace{7.0cm}
\includegraphics{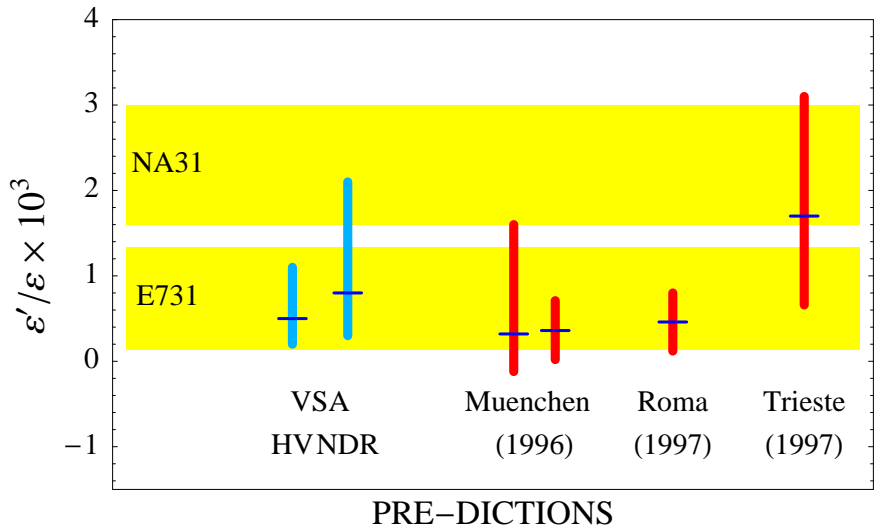}
\caption{
The 1-$\sigma$ results of the
NA31 and E731 Collaborations (early 90's)
are shown by the gray horizontal bands.
The old M\"unchen, Roma and Trieste theoretical predictions for \eoe are
depicted by the vertical bars with their central values.
For comparison, the VSA estimate is shown using two renormalization schemes.
\label{fig:pre}}
\end{figure}

The status of the theoretical 
predictions and experimental data available before 
the KTeV announcement in February 1999
is summarized in Fig.~\ref{fig:pre}. 

The gray horizontal bands show
the one-sigma average of the old (early 90's)
NA31 (CERN) and E731 (Fermilab) results. 
The vertical lines show the ranges of the
published theoretical $predictions$ (before February 1999), 
identified with the cities where most of the group members reside.
The range of the naive Vacuum Saturation Approximation (VSA) is
shown for comparison. 

\begin{figure}[t]
\vspace{7.0cm}
\includegraphics{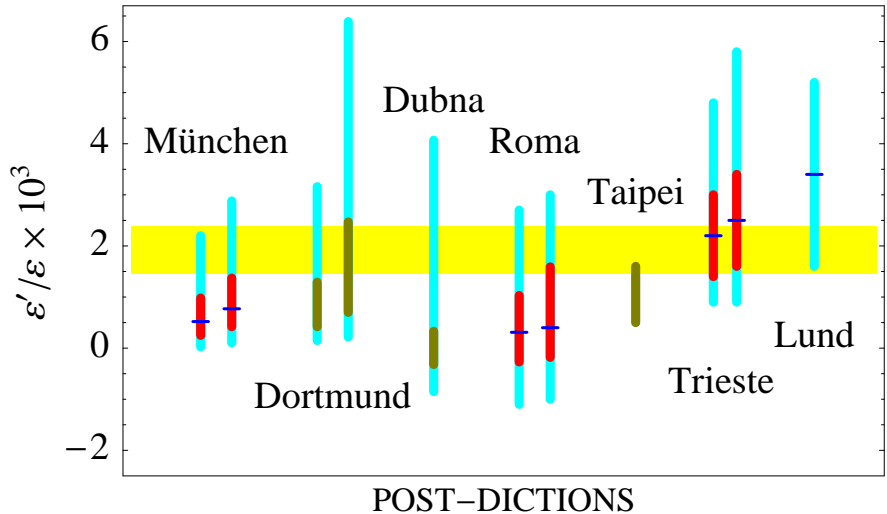}
\caption{ 
The latest theoretical calculations of \eoe are compared with
the combined 1-$\sigma$ average of the
NA31, E731, KTeV and NA48 results (\eoe = $19.2\pm 4.6\times 10^{-4}$),
depicted by the gray horizontal band (the error is inflated
according to the Particle Data Group procedure when averaging over
data with substantially different central values).  
\label{fig:post}}
\end{figure}

The experimental and theoretical scenarios have changed substantially
after the first KTeV data and the subsequent NA48 results.   
Fig.~\ref{fig:post} shows the present experimental world average
for \eoe compared with the revised or new theoretical calculations
that appeared during the last year. 

Notwithstanding the complexity of the problem,
all theoretical calculations 
show a remarkable overall agreement,
most of them pointing to a non-vanishing positive effect in the SM
(which is by itself far from trivial).

On the other hand, if we focus our attention on
the central values,
many of the predictions
prefer the $10^{-4}$ regime, whereas only a few of them stand 
above $10^{-3}$.
Is this just a ``noise'' in the theoretical calculations?

The answer is no.
Without entering the details of the various estimates, 
it is possible to explain most of
the abovementioned difference in terms
of a single effect: the different size of the 
hadronic matrix element of the gluonic penguin $Q_6$
obtained in the various approaches. 

While some of the calculations, as the early M\"unchen and Rome predictions,
assume for $\vev{\pi\pi|Q_6|K}$ values in the neighboroud of the 
leading $1/N$ result (naive factorization), other approaches,
among which the Trieste and Dortmund calculations,
find a substantial enhancement 
of this matrix element with respect to the simple factorization result.
The bulk of such an effect
is actually a global enhancement of the
$I=0$ components of the $K\to\pi\pi$ amplitudes, 
which affects $both$ current-current $and$ penguin operators, and
it can be clearly understood in terms of 
chiral dynamics (final-state interactions).

As a matter of fact, one should in general expect an enhancement of
\eoe with respect to the naive VSA due to final-state interactions (FSI). 
As Fermi first argued\cite{Fermi}, in potential scattering 
the isospin $I=0$ two-body~states feel an 
attractive interaction, of a sign opposite to that of the $I=2$
components thus affecting the size of the corresponding amplitudes. 
This feature is at the root of the enhancement of the
$I=0$ amplitude over the $I=2$ one and of the corresponding enhancement
of \eoe beyond factorization.

The question is how to make of a qualitative statement 
a quantitative one.
Recent dispersive studies of the $K\to\pi\pi$ amplitudes\cite{dispersive} 
make use of the Omn\`es-Mushkelishvili representation\cite{Omnes}
\be 
M(s+i\epsilon) = P(s)\ {\rm exp}\left(\frac{1}{\pi}
\int_{4 m_\pi^2}^\infty\frac{\delta(s')}{s'-s-i\epsilon} ds'\right)\ ,
\label{omnes}
\ee
in order to resum FSI effects from the knowledge of the
$\pi\pi$ rescattering phase $\delta(s)$ in the elastic 
regime ($s < 1$ GeV$^2$).
$P(s)$ is a polinomial function of $s$ which is related
to the factorized amplitude.
A solution of the above dispersive relation for the $A_{0,2}$
amplitudes can be written as
\be
A_I(s) = A'\ (s-m_\pi^2)\ {\cal R}_I(s)\ e^{i\delta_I(s)}\ ,
\label{omnes-sol}
\ee
where $A'$ is the derivative of the amplitude at the subtraction
point $s=m_\pi^2$. The coefficient $\cal R$ represents the 
rescaling effect related to the FSI. By replacing $A'_I$ with
the value given by LO chiral perturbation theory,
Pich and Pallante found ${\cal R}(m_k^2)_{0,2} \simeq  1.4,\ 0.9$
thus confirming via a (partial) resummation of FSI the enhancement
of the $I=0$ amplitudes, together with a mild depletion of
the $I=2$ components. 

The numerical significance of these results has been 
questioned\cite{burasetal} on the basis that
the precise size of the effect 
depends on the boundary conditions of the factorized amplitude which are
not unambiguously known, due to the neglect of 
higher order chiral corrections.
On the other hand, one should keep in mind that 
the choice of a low subtraction scale minimizes
the effect of chiral corrections and makes the result of the dispersive
analysis trustworthy.

At any rate a question remains: 
is the FSI rescaling on the factorized amplitudes, 
albeit a crucial contribution to the calculation of \eoe,
all what we were looking for ?

Given the possibility that common systematic uncertainties may 
affect the calculation of \eoe and the $\Delta I = 1/2$ rule
(see for instance the present difficulties in calculating on the lattice
``penguin contractions'' for CP violating as well as for
CP conserving amplitudes\cite{ciuchini})
a convincing calculation of \eoe must involve at the same time
a reliable explanation of the $\Delta I = 1/2$ selection rule,
which is still missing. FSI effects alone are {\em not} enough to account
for the large ratio of the $I=0,2$ CP conserving amplitudes.
Other sources of large non-factorizable corrections are needed, which
may affect the determination of \eoe as well. 

The $\Delta I = 1/2$ selection rule in $K\to\pi\pi$ decays is known since 
45 years\cite{Pais-Gell-Mann} and it states the experimental
evidence that kaons
are 400 times more likely to decay in the $I=0$ two-pion state
than in the $I=2$ component. This rule is not justified by any
general symmetry consideration and, although it is common understanding 
that its explanation must be rooted in the dynamics of strong interactions,
there is no up to date derivation of this effect from first principle QCD. 

The ratio of $I=2$ over $I=0$  amplitudes appears 
directly in the definition of \eoe:
\bea
\frac{\varepsilon'}{\varepsilon} &=& \frac{1}{\sqrt{2}} \left\{
\frac{\langle ( \pi \pi )_{I=2} | {\cal H}_W | {K_L} \rangle}
{\langle ( \pi \pi )_{I=0} | {\cal H}_W | {K_L} \rangle} \right.
 - \left.
\frac{\langle ( \pi \pi )_{I=2} | {\cal H}_W | {K_S} \rangle}
{\langle ( \pi \pi )_{I=0} | {\cal H}_W | {K_S} \rangle} \right\}\ .
\label{eedi}
\eea
As a consequence, a consistent calculation of \eoe must also address 
the determination of the CP conserving amplitudes. 

The way we approach the calculation of the hadronic $K\to\pi\pi$ transitions
in gauge theories is provided by the Operator Product Expansion 
which allows us to write
the relevant amplitudes in terms of the hadronic matrix elements of
effective $\Delta S = 1$ four quark operators (at a scale $\mu$)
and of the corresponding Wilson coefficients,
which encode the information about the dynamical degrees
of freedom heavier than the chosen renormalization scale:   
\be
{\cal H}_{\Delta S = 1} =
\frac{G_F}{\sqrt{2}} V_{ud}\,V^*_{us} \sum_i \Bigl[{z_i}(\mu) 
+ {\tau}\ {y_i}(\mu) \Bigr] {Q_i} (\mu) \ .
\label{Leff}
\ee
The entries $V_{ij}$ of the $3\times 3$ Cabibbo-Kobayashi-Maskawa
matrix describe the flavour mixing in the SM
and ${\tau} = - {V_{td}V_{ts}^{*}}/V_{ud}V_{us}^{*}$.
For $\mu < m_c$ ($q=u,d,s$), the relevant quark operators are:
\be
\begin{array}{lcl}
\left.
\begin{array}{lcl}
{Q_{1}} & = & \left( \overline{s}_{\alpha} u_{\beta}  \right)_{\rm V-A}
            \left( \overline{u}_{\beta}  d_{\alpha} \right)_{\rm V-A}
\\[1ex]
{Q_{2}} & = & \left( \overline{s} u \right)_{\rm V-A}
            \left( \overline{u} d \right)_{\rm V-A}
\end{array}
\right\} &&\hspace{-0em} \mbox{Current-Current}
\\[4ex]
\left.
\begin{array}{lcl}
{Q_{3,5}} & = & \left( \overline{s} d \right)_{\rm V-A}
   \sum_{q} \left( \overline{q} q \right)_{\rm V\mp A}
\\[1ex]
{Q_{4,6}} & = & \left( \overline{s}_{\alpha} d_{\beta}  \right)_{\rm V-A}
   \sum_{q} ( \overline{q}_{\beta}  q_{\alpha} )_{\rm V\mp A}
\end{array}
\right\} &&\hspace{-0em} \mbox{Gluon ``penguins''}
\\[4ex]
\left.
\begin{array}{lcl}
{Q_{7,9}} & = & \frac{3}{2} \left( \overline{s} d \right)_{\rm V-A}
         \sum_{q} \hat{e}_q \left( \overline{q} q \right)_{\rm V\pm A}
\\[1ex]
{Q_{8,10}} & = & \frac{3}{2} \left( \overline{s}_{\alpha} 
                                                 d_{\beta} \right)_{\rm V-A}
     \sum_{q} \hat{e}_q ( \overline{q}_{\beta}  q_{\alpha})_{\rm V\pm A}
\end{array}
\right\} &&\hspace{-0em} \mbox{Electroweak ``penguins''}
\end{array} 
\label{quarkeff}
\ee
Current-current operators are induced by tree-level W-exchange whereas
the so-called penguin (and ``box'') diagrams are generated via an 
electroweak loop.
Only the latter ``feel'' all three quark families via the virtual quark
exchange and are therefore sensitive to the weak CP phase.
Current-current operators control instead the CP conserving
transitions. This fact suggests already that the connection 
between \eoe and the
$\Delta I = 1/2$ rule is by no means a straightforward one.

Using the effective $\Delta S=1$ quark Hamiltonian we can write \eoe as
\be
\frac{{\varepsilon'}}{\varepsilon} =  
e^{i \phi} \frac{G_F \omega}{2|\epsilon|\Re{A_0}} \:
{\mbox{Im}\, \lambda_t} \: \:
 \left[ {\Pi_0} - \frac{1}{\omega} \: {\Pi_2} \right]
\label{main}
\ee
where
\be
\begin{array}{lcl}
 {\Pi_0} & = & \frac{1}{{\cos\delta_0}} 
\sum_i {y_i} \, 
\Re\langle  Q_i  \rangle _0\ (1 - {\Omega_{\eta +\eta'}}) 
\\[1ex]
 {\Pi_2} & = & \frac{1}{{\cos\delta_2}} \sum_i {y_i} \, 
\Re\langle Q_i \rangle_2 \quad ,
\end{array} 
\label{PI02}
\ee
and $\langle Q_i \rangle \equiv \langle \pi\pi | Q_i | K \rangle$.
The rescattering phases $\delta_{0,2}$ can be extracted from elastic $\pi-\pi$
scattering data\cite{FSIphases} and are such that $\cos\delta_0 \simeq 0.8$
and $\cos\delta_2 \simeq 1$. Given that the phase of $\varepsilon$,
$\theta_\varepsilon$, is approximately $\pi/4$, 
as well as $\delta_0-\delta_2$, 
$\phi = \frac{\pi}{2} + {\delta_2} - {\delta_0} - \theta_\varepsilon$ 
turns out to be consistent with zero.

Two key ingredients appear in eq. \ref{main}:
\begin{itemize}
\item[1.]
The isospin breaking $\pi^0-\eta-\eta'$ mixing, parametrized by
$\Omega_{\eta+\eta'}$, which is estimated to give
a $positive$ correction to the $A_2$ amplitude of about 15-35\%.
However, it has been recently emphasized\cite{GardnerValencia} that
NLO chiral corrections may make $\Omega_{\eta+\eta'}$ negative and 
as large as $-0.7$, thus potentially providing
a strong enhancement mechanism for \eoe.
Further isospin breaking effects have been very recently 
discussed\cite{DelI=5/2}.

\item[2.]
The combination of CKM elements  
{$\Im \lambda_t \equiv \Im (V_{ts}^*V_{td})$}, which
affects directly the size of \eoe and
the range of the uncertainty.
The determination of $\Im \lambda_t$ goes through B-physics
constraints and $\varepsilon$.
The latter observable depends on the 
theoretical determination of $B_K$, the $\bar K^0-K^0$ hadronic parameter,
which should be consistently determined within every approach.
The theoretical uncertainty on $B_K$ affects subtantially the
final uncertainty on $\Im \lambda_t$.
A better determination of the unitarity triangle
is expected from the B-factories and the hadronic colliders\cite{CPprospects}.
From K-physics, {$K_L\to\pi^0\nu\bar\nu$} gives the cleanest ``theoretical''
determination of $\Im\lambda_t$, albeit representing a great experimental
challenge\cite{rareK}.
\end{itemize}

A satisfactory approach to the
calculation of \eoe should comply with the following requirements: 
\begin{itemize}
\item[A:]
A consistent definition of renormalized operators leading to
the correct scheme and scale matching with the short-distance
perturbative analysis.\\[-3ex]
\item[B:]
A self-contained calculation of $all$ relevant hadronic matrix elements
(including $B_K$).\\[-3ex]
\item[C:]
A simultaneous explanation of the $\Delta I = 1/2$ selection rule
and \eoe.
\end{itemize}

None of the available calculations satisfies all previous 
requirements. I summarize very briefly the various attempts
to calculate \eoe which have appeared so far and have lead to the
estimates shown in Figs. \ref{fig:pre}
and \ref{fig:post}.
A simple naive approach to the problem is
the {VSA}, which is based on two drastic assumptions:
the factorization of the four quark operators
and the vacuum saturation of 
the completeness of the intermediate states. 
As an example: 
\bea
\langle \pi^+ \pi^-|Q_6| K^0 \rangle & = & 
 2\  \langle \pi^-|\overline{u}\gamma_5 d|0 \rangle
\langle \pi^+|\overline{s} u |K^0 \rangle 
- 2\  \langle \pi^+ \pi^-|\overline{d} d|0 \rangle  
\langle 0|\overline{s} \gamma_5 d |K^0 \rangle
\nonumber \\
& & +\ 2  \left[\langle 0|\overline{s} s|0 \rangle - 
\langle 0|\overline{d}d|0 \rangle\right] 
\langle \pi^+ \pi^-|\overline{s}\gamma_5 d |K^0 \rangle
\eea
The VSA does not allow
for a consistent matching of the scale and scheme dependence
of the Wilson coefficients (the HV and NDR results are shown in
Fig. \ref{fig:pre}) and it carries a potentially
large systematic uncertainty\cite{review}. 
It is best used for LO estimates. 

Generalized Factorization represents an attempt to address
the issue of scale and renormalization scheme dependence 
in the framework of factorization 
by defining effective Wilson
coefficients which absorb the perturbative QCD running
of the quark operators. The new coefficients,
scale and scheme independent at a given order in the
strong coupling expansion, are matched to
the factorized matrix elements at the scale {$\mu_F$}
which is arbitrarily chosen in the perturbative
regime. A residual
scale dependence remains in the penguin matrix elements via the
quark mass.
Fitting the $\Delta I = 1/2$ rule and \eoe requires non-factorizable
contributions both in the current-current
and the penguin matrix elements, parametrized 
by (independent) phenomenological parameters\cite{Cheng}.

In the M\"unchen approach 
(phenomenological $1/N$) some of the matrix elements are obtained by
fitting the $\Delta I =1/2$ rule at $\mu=m_c=1.3$ GeV.
On the other hand, the relevant gluonic and electroweak penguin 
$\vev{Q_6}$ and $\vev{Q_8}_2$ remain undetermined and 
are taken around their leading $1/N$ values 
(which implies a scheme dependent result).
In Fig. \ref{fig:post} the HV (left) and NDR (right) results
are shown\cite{Buras}. The dark range represents the result of gaussian
treatment of the input parameters compared to flat scannning 
(complete range). 

In the recent years the Dortmund group has revived and improved
the approach of Bardeen, Buras and Gerard\cite{BBG} based on the $1/N$
expansion. 
Chiral loops are regularized via a cutoff 
and the amplitudes are arranged in a $p^{2n}/N$ expansion. 
A particular attention has been given
to the matching procedure between the scale dependence of the chiral loops
and that arising from the short-distance analysis\cite{Dortmund}.  
The renormalization scheme dependence remains and it is included in the 
final uncertainty. The $\Delta I = 1/2$ rule is reproduced, but the
presence of the quadratic cutoff induces a matching scale instability
(which is very large for $B_K$).
The NLO corrections to $\vev{Q_6}$ induce a substantial enhancement
of the matrix element (right range in Fig. \ref{fig:post}) 
compared to the leading order result (left).
The dark range is drawn for central values of
$m_s$, $\Omega_{\eta+\eta'}$, $\Im \lambda_t$ and $\Lambda_{QCD}$. 

In the Nambu, Jona-Lasinio (NJL) modelling of QCD\cite{NJL} 
the Dubna group\cite{Belkov}
has calculated \eoe including
chiral loops up to $O(p^6)$ and the effects of 
scalar, vector and axial-vector resonances.
Chiral loops are regularized via
the heat-kernel method, which leaves unsolved the
problem of the renormalization scheme dependence.
A phenomenological fit of the $\Delta I = 1/2$ rule implies deviations up to 
a factor two on the calculated $\vev{Q_6}$.  
The reduced (dark) range in Fig. \ref{fig:post} corresponds to taking the   
central values of the NLO chiral couplings and varying the short-distance 
parameters.

In the approach of the Trieste group, based on the
Chiral Quark Model ($\chi$QM)\cite{chiQM}, 
all hadronic matrix elements are computed up to $O(p^4)$ in the chiral 
expansion in terms of the three model parameters:
the constituent quark mass, the quark condensate
and the gluon condensate.
These parameters are
phenomenologically fixed by fitting the $\Delta I =1/2$ rule\cite{ts98a}.
This step is crucial in order to make the model predictive, 
since there is no a-priori argument
for the consistency of the matching procedure. 
As a matter of fact, all computed observables turn
out to be very weakly dependent on the
scale (and the renormalization scheme)
in a few hundred MeV range around the 
matching scale, which is taken at $0.8$ GeV as a
compromise between the ranges of validity 
of perturbation theory and chiral expansion.
The $I=0$ matrix elements are strongly enhanced by non-factorizable
contributions and drive \eoe in the $10^{-3}$ regime.
The dark (light) ranges in Fig. \ref{fig:post} correspond 
to Gaussian (flat) scan of the input parameters. 
The bars on the left represent the results
of ref. \cite{ts00a} which updates the 1997 calculation\cite{ts98b}.
That on the right are a new estimate\cite{ts00b}, similarly 
based on the $\chi$QM hadronic matrix elements, in which however
\ratio\ is estimated in a novel way
by including the  explicit computation of \eps\ in the ratio as
opposed to  the usual procedure of taking its value from the experiments. 
This approach has the advantage of
being independent from the determination of the 
CKM parameters $\Im \lambda_t$ and of showing
more directly the dependence on the long-distance parameter $\hat B_K$
as determined within the model.

Lattice regularization of QCD
is $the$ consistent approach to the problem. 
On the other hand, there are presently important numerical and 
theoretical limitations, like the quenching approximation
and chiral symmetry, which 
may substantially affect the calculation of the weak matrix elements. 
In addition, chiral perturbation
theory is needed in order to obtain $K\to\pi\pi$ amplitudes  
from the computed $K\to\pi$ transitions.
As summarized by Ciuchini at this conference\cite{ciuchini}
lattice cannot provide us at present with reliable calculations
of the $I=0$ penguin operators relevant to \eoe, as well as of the 
$I=0$ components of the hadronic matrix elements of the
current-current operators (penguin contractions), which are relevant to the
$\Delta I = 1/2$ rule. 
This is due to large renormalization uncertainties,
partly related to the breaking of chiral symmetry on the lattice.
In this respect, very promising is the Domain Wall Fermion
approach\cite{DWF} which allows us to decouple the chiral symmetry 
from the continuum limit\cite{Soni}.
In the recent Roma re-evaluation of \eoe 
$\vev{Q_6}$ is taken at the VSA value with a 100\% uncertainty\cite{Roma}.
The result is therefore scheme dependent (the HV and NDR results
are shown in Fig. \ref{fig:post}). The dark (light) ranges correspond 
to Gaussian (flat) scan of the input parameters. 

The $\Delta I = 1/2$ rule and $B_K$ have been 
studied in the NJL framework and $1/N$ expansion
by Bijnens and Prades\cite{BijnensPrades} showing an impressive scale 
stability when including vector and axial-vector resonances.
The same authors have
recently produced a calculation of \ratio\ at the 
NLO in $1/N$\cite{lund}. The calculation is 
done in the chiral limit and it is eventually corrected 
by estimating the largest $SU(3)$ breaking effects. 
Particular attention is devoted to the matching between
long- and short-distance components by use of the $X$-boson 
method\cite{X-boson0,X-boson1}. The couplings of the $X$-bosons are computed
within the ENJL model which improves the high-energy behavior.
The $\Delta I = 1/2$ rule is reproduced and the computed amplitudes show a 
satisfactory renormalization scale and scheme stability. 
A sizeable enhancement of the
$Q_6$ matrix element is found which brings the central value of \ratio
at the level of $3\times 10^{-3}$.

Other attempts to reproduce the measured \eoe using the linear 
$\sigma$-model, which include the effect of a scalar resonance 
with $m_\sigma \simeq 900$ MeV, obtain the needed enhancement
of $\vev{Q_6}$\cite{sigmamodel}. However, it is not possible
to reproduce simultaneously the experimental values of \eoe and of
the CP conserving $K\to\pi\pi$ amplitudes. 

Studies on the matching between long- and short- distances in large
$N$ QCD, with the calculation of the $Q_7$ penguin matrix element
and of $\hat B_K$ at the NLO in the $1/N$ expansion
have been presented in ref.\cite{PerisEdR}. However,
a complete calculation of the $K\to\pi\pi$ matrix elements
relevant to \eoe is not available yet.

\begin{figure}
\vspace{7.0cm}
\includegraphics{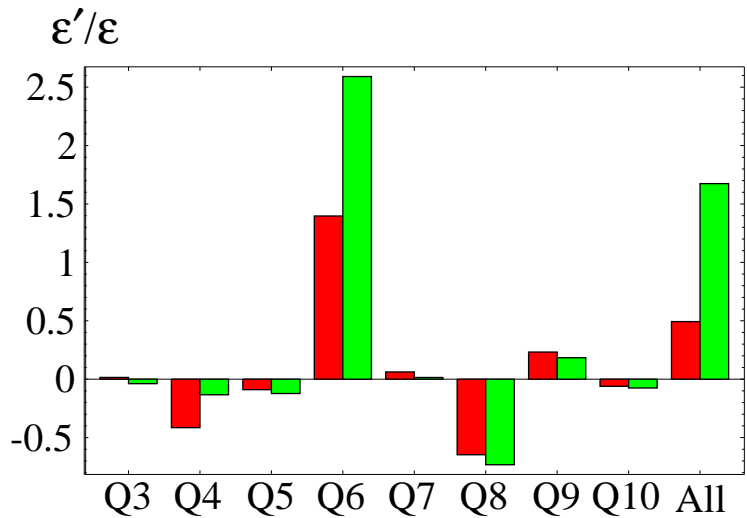}
\caption{Predicting $\varepsilon'/\varepsilon$:
a (Penguin) Comparative Anatomy of the M\"unchen (dark gray) and 
Trieste (light gray) results in Fig. \ref{fig:pre} (in units of $10^{-3}$).
\label{istobuqm}}
\end{figure}

Without entering into the details of the various calculations
I wish to illustrate with a simple exercise the crucial
role of final-state interactions (in general of 
non-factorizable contributions) for \eoe and the $\Delta I = 1/2$ 
selection rule.
In order to do that I focus on two semi-phenomenological approaches.

A commonly used way of comparing the estimates of hadronic matrix elements
in different approaches is via the so-called $B$ factors which
represent the ratio of the model matrix elements to the corresponding VSA
values. However, care must be taken in the comparison of
different models due to the scale
dependence of the $B$'s and the values used by different groups
for the parameters that enter the VSA expressions. 
An alternative pictorial and synthetic
way of analyzing different outcomes for \eoe
is shown in Fig.~\ref{istobuqm}, where a ``comparative
anatomy'' of the old Trieste and M\"unchen predictions
is presented. 

From the inspection of the various contributions it is apparent that
the different outcome on the central value of \eoe is almost entirely
due to the difference in the $Q_6$ component. 

In the M\"unich approach\cite{Buras} the $\Delta I = 1/2$ rule
is used in order to determine phenomenologically the  
matrix elements of $Q_{1,2}$ and, 
via operatorial relations, some of the matrix elements of the
left-handed penguins. The approach does not allow
for a phenomenological determination of the matrix elements of the penguin
operators which are most relevant for \eoe, namely the gluonic penguin $Q_6$
and the electroweak penguin $Q_8$. These matrix elements are taken
around their leading $1/N$ values (factorization). 

In the semi-phenomenological approach of the Trieste group 
the size of the effects on the $I=0,2$ amplitudes is controlled
by the phenomenological embedding of the $\Delta I= 1/2$ selection rule
which determines the ranges of the model paremeters: 
the constituent quark mass, the quark and the gluon condensates.
In terms of these parameters all matrix elements are computed.

\begin{figure}
\vspace{7.0cm}
\includegraphics{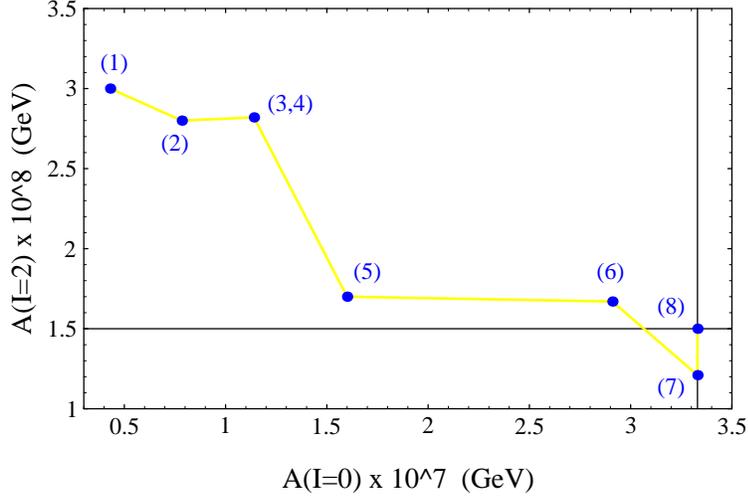}
\caption{Anatomy of the $\Delta I = 1/2$ rule in the 
$\chi$QM. See the text for explanations.
The cross-hairs indicate the experimental point. 
\label{road}}
\end{figure}

Fig.~\ref{road} shows an anatomy of the $\chi$QM contributions
which lead to the experimental value
of the $\Delta I = 1/2$ selection rule for central values of the 
input parameters.
Point (1) represents the result obtained by neglecting QCD
and taking the factorized matrix element for the
tree-level operator $Q_2$, which is the leading electroweak 
contribution. 
The ratio $A_0/A_2$ is thus found equal
to $\sqrt{2}$: by far off the experimental point (8).
Step (2) includes the effects of perturbative QCD renormalization
on the operators $Q_{1,2}$\cite{Gaillard-etc}. 
Step (3) shows the effect of including the gluonic 
penguin operators\cite{VSZ-GWise-CFGeorgi}. 
Electroweak penguins\cite{Lusignoli-etc} are numerically
negligeable in the CP conserving amplitudes and 
are responsible for the very small shift in the $A_2$ direction.
Therefore, perturbative QCD and factorization lead us from (1) to (4).

Non-factorizable gluon-condensate corrections,
a crucial model dependent effect entering at the leading order
in the chiral expansion, produce a substantial
reduction of the $A_2$ amplitude (5), 
as it was first observed by Pich and de Rafael\cite{Pich-deRafael}. 
Moving the analysis to $O(p^4)$,
the chiral loop corrections, computed on the LO chiral
lagrangian via dimensional regularization and minimal subtraction, 
lead us from (5) to (6), while 
the finite parts of the NLO counterterms
calculated in the $\chi$QM approach lead to the point (7).
Finally, step (8) represents the inclusion of $\pi$-$\eta$-$\eta'$ 
isospin breaking effects\cite{Ometapeta}. 

This model dependent anatomy 
shows the relevance of non-factorizable contributions
and higher-order chiral corrections. The suggestion that
chiral dynamics may be relevant to
the understanding of the $\Delta I = 1/2$ selection rule goes back to the
work of Bardeen, Buras and Gerard\cite{BBG} in the $1/N$
framework with a cutoff regularization. 
A pattern similar to that shown
in Fig.~\ref{road} for the chiral loop corrections to $A_0$ and $A_2$
was previously obtained in a NLO chiral
lagrangian analysis, using dimensional regularization, by 
Missimer, Kambor and Wyler\cite{MKWyler}.
The Trieste group has extended their calculation to include
the NLO contributions
to the electroweak penguin matrix elements\cite{ts96a,ts98b}.

\begin{figure}
\vspace{7.0cm}
\includegraphics{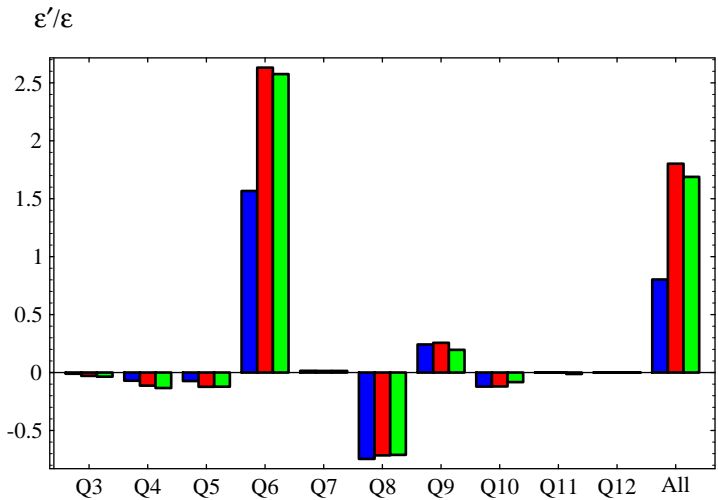}
\caption{Anatomy of \eoe (in units of $10^{-3}$) in the $\chi$QM 
approach. 
In black the LO results (which includes the non-factorizable gluonic
corrections), in half-tone the effect of the inclusion of chiral-loop 
corrections and in light gray the complete $O(p^4)$ estimate.
\label{charteps}}
\end{figure}

Fig.~\ref{charteps} shows the contributions to \eoe
of the various penguin operators, providing us
with a finer anatomy of the NLO chiral corrections.
It is clear that chiral-loop dynamics plays a subleading role in the 
electroweak penguin sector ($Q_{8-10}$) while enhancing by 60\% the gluonic
penguin ($I=0$) matrix elements. 
The NLO enhancement of the $Q_6$ matrix element is what drives \eoe 
in the $\chi$QM to the $10^{-3}$ ballpark.

As a consequence, the $\chi$QM analysis shows that the same dynamics 
that is relevant to the
reproduction of the CP conserving $A_0$ amplitude 
(Fig.~\ref{road}) is at work
in the CP violating sector, albeit with a reduced strenght. 

In order to ascertain
whether these model features represent real QCD effects we must
wait for future improvements in lattice calculations\cite{ciuchini,Soni}.
On the other hand, indications for such a dynamics follow also from
the analitic properties
of the $K\to\pi\pi$ amplitudes, as the dispersive analyses
show\cite{dispersive}.
It is important to notice however that the size of
the effect so derived is generally
not enough to fully account for the $\Delta I = 1/2$ rule.
Other non-factorizable contributions are needed to further
enhance the CP conserving $I=0$ amplitude, and to
reduce the large $I=2$ amplitude obtained from perturbative QCD
and factorization.
In the $\chi$QM approach, for instance, the fit of the  $\Delta I = 1/2$ rule
is due to the interplay of FSI (at the NLO) and non-factorizable soft gluonic
corrections (at LO in the chiral expansion). 

The idea of a connection between the $\Delta I = 1/2$ selection
rule and \eoe goes back a long way\cite{VSZacharov&Gilman-Wise},
although at the GeV scale, where we can trust perturbative
QCD, penguins are far from providing
the dominant contribution to the CP conserving amplitudes.

In summary,
those semi-phenomenological approaches which reproduce the
$\Delta I = 1/2$ selection rule in $K\to\pi\pi$ decays,
generally agree in the pattern and size    
of the $I = 2$ hadronic matrix elements with the existing
lattice calculations. 
On the other hand,
the $\Delta I = 1/2$ rule forces upon us large deviations from
the naive VSA for the $I=0$ amplitudes: 
$B-$factors of $O(10)$ are required for $\vev{Q_{1,2}}_0$.
Here is were 
lattice calculations suffer from large sistematic uncertainties. 

In the Trieste and Dortmund calculations, which reproduce the CP conserving
$K\to\pi\pi$ amplitudes, non-factorizable effects
(mainly final-state interactions) enhance the hadronic matrix 
elements of the gluonic penguins, and give $B_6/B_8^{(2)} \approx 2$. 
Similar indications stem from recent $1/N$\cite{lund} 
and dispersive\cite{dispersive} approaches. 
The direct calculation of $K\to\pi\pi$ amplitudes and unquenching
are needed in the lattice calculations in order to account
for final state interactions. Further progress in this direction
is awaited.


%
\end{document}